\documentclass[fleqn,twoside]{article}
\usepackage{espcrc2}

\usepackage{graphicx}
\usepackage[figuresright]{rotating}

\newcommand{\AmS}{{\protect\the\textfont2
  A\kern-.1667em\lower.5ex\hbox{M}\kern-.125emS}}

\title{UHE Cosmic Rays and Neutrinos Showering on Planet
Edges
  }

\author{D. Fargion ,
        P. Oliva, O. Lanciano,  \\ {Physics Department and INFN,
        Universita'  La Sapienza, Roma, Ple. A. Moro 2,00185, Italy}%
        }

\begin{document}

\begin{abstract}
Ultra High Energy  (UHE) Cosmic Rays , UHECR, may graze high
altitude atmosphere leading to horizontal upward air-showers. Also
PeVs $\bar{{\nu_e}}$ hitting electron in atmosphere may air-shower
at $W^-$ boson resonant mass.  On the other side ultra high energy
muon and electron neutrinos  may also lead, by UHE neutrinos mass
state mixing, to the rise of a corresponding UHE Tau neutrino
flavor; the consequent UHE tau neutrinos, via charge current
interactions in matter, may create UHE taus at horizons (Earth
skimming neutrinos or Hor-taus) whose escape in atmosphere and
whose consequent decay in flight, may be later amplified by upward
showering on terrestrial, planetary atmospheres. Indeed because of
the finite terrestrial radius $R_{\oplus}$, its thin atmosphere
size ($h_0\simeq10$) km, its dense crust, the UHE tau cannot
extend much more than $\sim\sqrt{R_{\oplus}h_0\pi/2}\simeq360$
kilometers in air, corresponding to an energy $E_{\tau
\oplus}\simeq7.2$ EeV, near but below  GZK cut-off ones; on the
contrary Jupiter (or even Saturn) may offer a wider, less dense
and thicker gaseous layer at the horizons where Tau may loose
little energy, travel longer before decay and  rise and shower  at
$E_{\nu_{\tau}}\simeq4-6\cdot10^{19}$ eV or ZeV extreme energy.
Also solar atmosphere may play a role, but unfortunately
tau-showers secondaries  maybe are too noisy to be disentangled,
while Jupiter atmosphere, or better, Saturn one, may offer a
clearer imprint for GZK (and higher Z-Burst) Tau showering, well
\emph{below } the horizons edges. \vspace{1pc}
\end{abstract}

\maketitle
\section{Past Neutrino Telescope Underground}
UHE Neutrino detection and Astronomy is a compelling cornerstone
in Astrophysics, possibly correlated to UHECR Astronomy
\cite{Grieder01}, just nearby GZK cut-off \cite{za66}. Most of
past and present $\nu$ detector are tracing muon in underground
detectors \cite{bh00}, \cite{Anchordoqui}. This because of
proliferous, penetrating nature of the muon. Due to the neutrino
flux paucity and its low cross-section most of the detector are
large as Super Kamiokande or larger volumes as AMANDA, Baikal or
Antares ones, looking to $km^3$ future size \cite{Gandhi98}.
Because most of cosmic rays are converting into atmospheric muon
secondaries, the muon dawn ward directions are polluted by
atmospheric noises. The up-ward muons, induced by UHE neutrinos
charged currents are suppressed because of their parental
neutrinos opacity to Earth sizes, above $40$ TeV energy range.
This energy unfortunately lays where astrophysical neutrinos might
dominate over atmospheric ones. Horizontal Muons might be a better
tools for UHE $\nu$ astronomy, but most km$^3$ detectors are
basically vertical string unable to disentangle  horizon
directions. However a novel way to observe UHE neutrinos has been
proposed since almost a decade \cite{Fargion1999} and revived on
recent years \cite{Fargion 2002a}, \cite{Bertou2002},
\cite{Feng2002}, \cite{Cao}, \cite{Tseng03}, \cite{Fargion2001},
\cite{Fargion2005} and \cite{Miele et. all05}; it is based on
horizontal skimming neutrinos whose consequent UHE $\tau$
air-showering by  decay in flight in air invade large detection
areas. The energy windows for Tau Astronomy opens at PeV up to few
EeV energy band which is of great interest.
\begin{figure}[t]
\includegraphics[width=65mm]{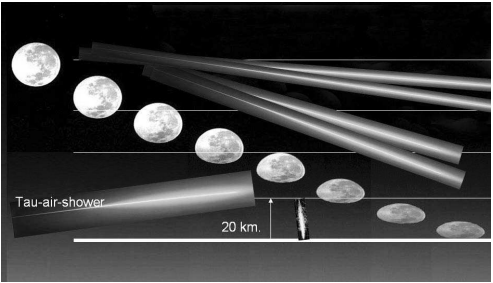}
\caption{\footnotesize{Long High Altitude Horizontal Air-showers
versus vertical ones: here we assumed a PeV-EeV C.R. and we
schematically draw the narrower beaming and the geomagnetic
splitting of the showers; the moon size from the Space Station at
Horizons is used as a ruler. A very rare (neutrino induced) Tau
air Shower may point upward within a noisy free Earth shadows,
coming from below the horizons }} \label{fig1}
\end{figure}
\begin{figure}[t] 
\includegraphics[width=70mm]{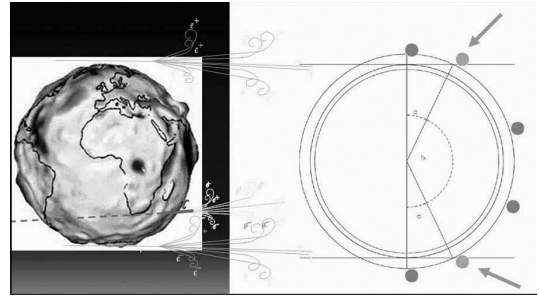}
\caption{\footnotesize{Showering at the Edges: The Atmosphere
UHECR are skimming the terrestrial atmosphere; a persistent gamma
source (as a AGN or BL Lac) at tens GeV-TeV or higher energies may
graze the air and bremsstrahlung   in lower X-ray cone dressed
around the Earth. The X-ray satellite might intersect this X-ray
cylinder twice in its keplerian orbit, revealing the hard $\gamma$
source in lower X-$\gamma$ band also with a re-brightening or
re-appearance of the source (\textit{shown by small arrows})
behind the Earth, depending on the orbit geometry; the phenomena
must be linked to the characteristic satellite period. Appearance
and reappearance might occur, for BATSE-like satellite, every
$93$, $36$, $57$ minute lag. The persistent Cherenkov beaming at
high altitudes (nearly thirty $km$ height) for horizontal gamma
showering has a narrow  angle $\simeq \pm 0.15^o$ and it may blaze
the satellite on Space Station altitudes as long as $\simeq 30 s$;
more rare upward Tau air-shower may blaze below the Earth edges.}}
\label{fig2}
\end{figure}
\begin{figure}[t]
\includegraphics[width=75mm]{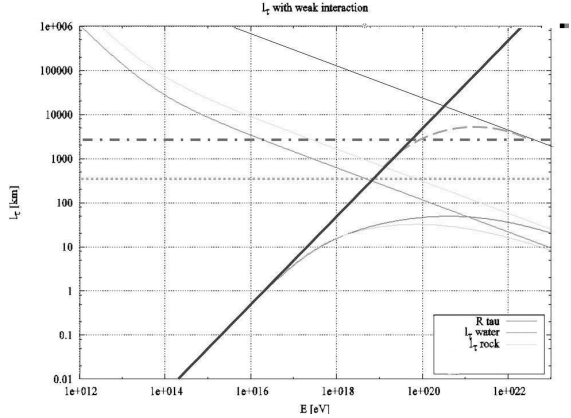}
 \caption  {\footnotesize{Tau length propagation $l_{\tau}$ in Terrestrial and Saturnian atmosphere.
 The  Tau linear growth with energy in vacuum (thick line) is
 compared with Tau propagation in terrestrial water (higher curve) and rock
(lowest curve) as well as in Jupiter or  Saturnian atmosphere
(dashed curve). In comparison the  line for horizons atmosphere on
Earth (dotted) and  for Saturn (dot-dashed). The UHE neutrino cord
penetrability in water or rock Earth do not exceed a few  $km$
depth, while on gaseous diluted Jupiter and Saturn planets the
cord trajectory may cross deeper (few hundreds) $km$ gas layers
travelling thousand kilometers with negligible absorption. The
arrival angle may be (depending on satellite orbit) better
disentangled from horizontal UHECR}}
\label{fig3}
\end{figure}
\begin{figure}[t]
\includegraphics[width=60mm]{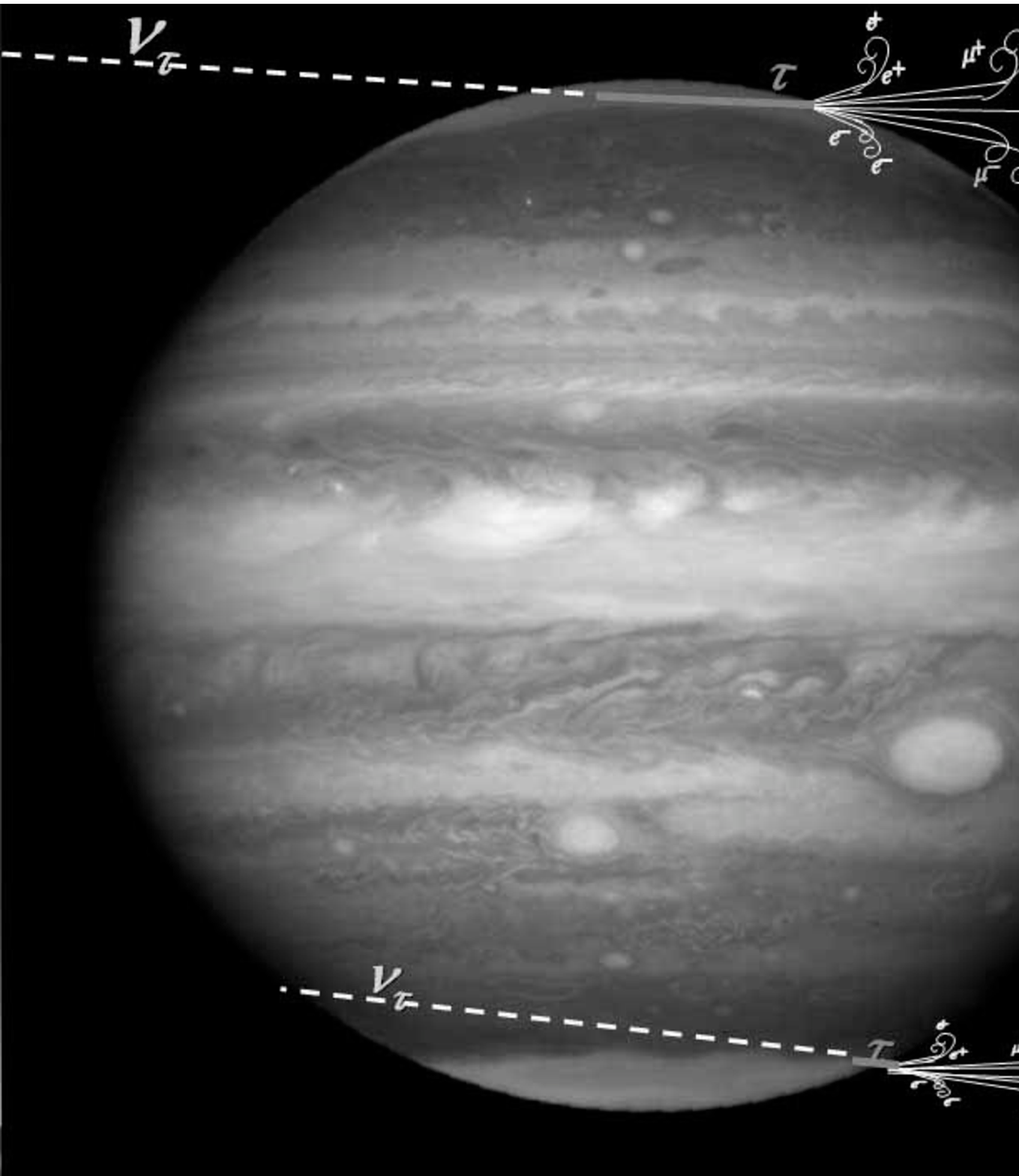}
\caption{\footnotesize{Schematic Showering in atmosphere  Jupiter
edges by $\nu_{\tau}$ charged current neutrino interaction and its
consequent ${\tau}$ decay in flight. The large planet radius makes
these ${\tau}$ flights much longer than terrestrial ones.}}
\label{fig4}
\end{figure}
\section{Atmosphere sizes versus $\tau$ tracks-energy}
Either UHE neutrino skimming showering in air  or up-going UHE tau
neutrino interacting within Earth, may be  leading to skimming
parent $\tau$ well amplified by a decay and a shower in flight.
The phenomena of such $\tau$ air-showers remind the double-bang
signature \cite{Learned Pakvasa 1995} for $\nu_{\tau}$ interaction
and soon later $\tau$ decay in water. Here we assume one bang
\textit{in} and the second bang \textit{out}. The simplest case of
a mountain Chain (like Ande) acting as a target for UHE neutrino
has been first considered \cite{Fargion1999}, while later the
Earth itself has been proposed as a beam dump at its edges
\cite{Fargion 2002a}, \cite{Fargion2004}. Their signal is
amplified and spread in huge number of secondaries over wide
areas. The $\nu_{\tau}\rightarrow\tau $ conversion into observable
Tau Shower may reach high efficiency at the horizontal atmosphere
edges, otherwise (in upward vertical directions)  their parental
UHE neutrinos suffer of severe planet opacity. The EeVs UHE tau
shower secondary (UV-X-Gamma-Muon Bundle) discover from Earth is
already at hand either by top mountains telescope as Magic ones,
or by sharp anisotropy in Horizontal Showers within Auger Ande
shadows, or by present and future satellites (like Swift, Glast
and future ideal Array detectors in Space Station, see Fig.
\ref{fig15} facing the Earth); these showering reminds the
eventual PeVs $\bar{{\nu_e}}$ hitting electron at $W^-$ boson
resonant mass. In analogy hypothetical UHE SUSY neutralino may be
scattering atmosphere electron at tens PeVs resonant energy for
$\tilde{e}$ s-channel, arriving \emph{above} the horizons
\cite{Datta}.
  At higher altitudes UHECR may skim the terrestrial or other
planetary atmosphere leading to thin collimated air-showers whose
structure is not just like conical  vertical ones, see Fig.
\ref{fig1}, but they often split in twin  jets   by geomagnetic
fields \cite{Fargion2001}. A simple analogous gamma originated
grazing mechanism may play a role by converting unobservable hard
GeV-TeV or PeVs gamma astronomy: while such hard photons are
skimming highest quota they are showering in terrestrial,
planetary or solar atmosphere altitudes; their skimming
secondaries produce a cylindric cone that may be intersect by a
satellite trajectory. These \textit{rises and dawns} at the
horizons of hard source  into softer one, might lead to transient
X sources whose set and re-appearance time  may exhibit
characteristic terrestrial (orbital) lag modulation see Fig.
\ref{fig2}.

While the UHE neutrino skimming the Earth produces an escaping
tau, the consequent decay in flight is constrained by the size of
the planet gas layers: on Earth the height size ($h_{0\oplus}$)
imply a maximal distance : $d_{\tau\oplus}\simeq\sqrt{R_{\oplus}
h_0\pi/2}\simeq360$ km. The corresponding maximal tau energy is
$E_{\tau\oplus}\simeq[d_{\tau\oplus}/(c\tau_{\tau})]m_{\tau}c^2\simeq7.2$
EeV. To visualize Horizontal Air-Showers versus vertical ones and
upward Hortau  see Fig. \ref{fig2}.

The larger the planet size $R$ and the wider its density height
scale  $h_0$, the larger is the allowed distance  $d_{\tau}$ for
the Tau to fly before  decay (and the higher its correlated
energy). The atmosphere and the planet density profile  rule also
the penetrability of the UHE primary neutrino. The case for the
Earth has been carefully analyzed in recent papers
(\cite{Fargion2004} and \cite{Fargion2004b}) and will be reminded
later.

\begin{figure}[t]
\includegraphics[width=70mm]{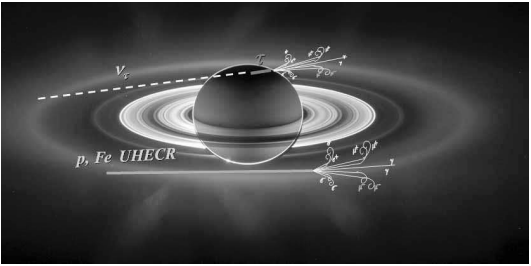}
\caption{\footnotesize{Schematic Showering in atmosphere (for
Ultra High Energy Cosmic Rays) and inside the Saturn edges by
$\nu_{\tau}$ charged current neutrino interaction and its
consequent ${\tau}$ decay in flight. The large planet radius and
its longest atmosphere height size makes these ${\tau}$ flights
the longest ones observable  in our planetary system.  Also very
peculiar UHECR showering takes place also along the thin Saturn
disk edges }} \label{fig5}
\end{figure}
\begin{figure}[t]
\includegraphics[width=70mm]{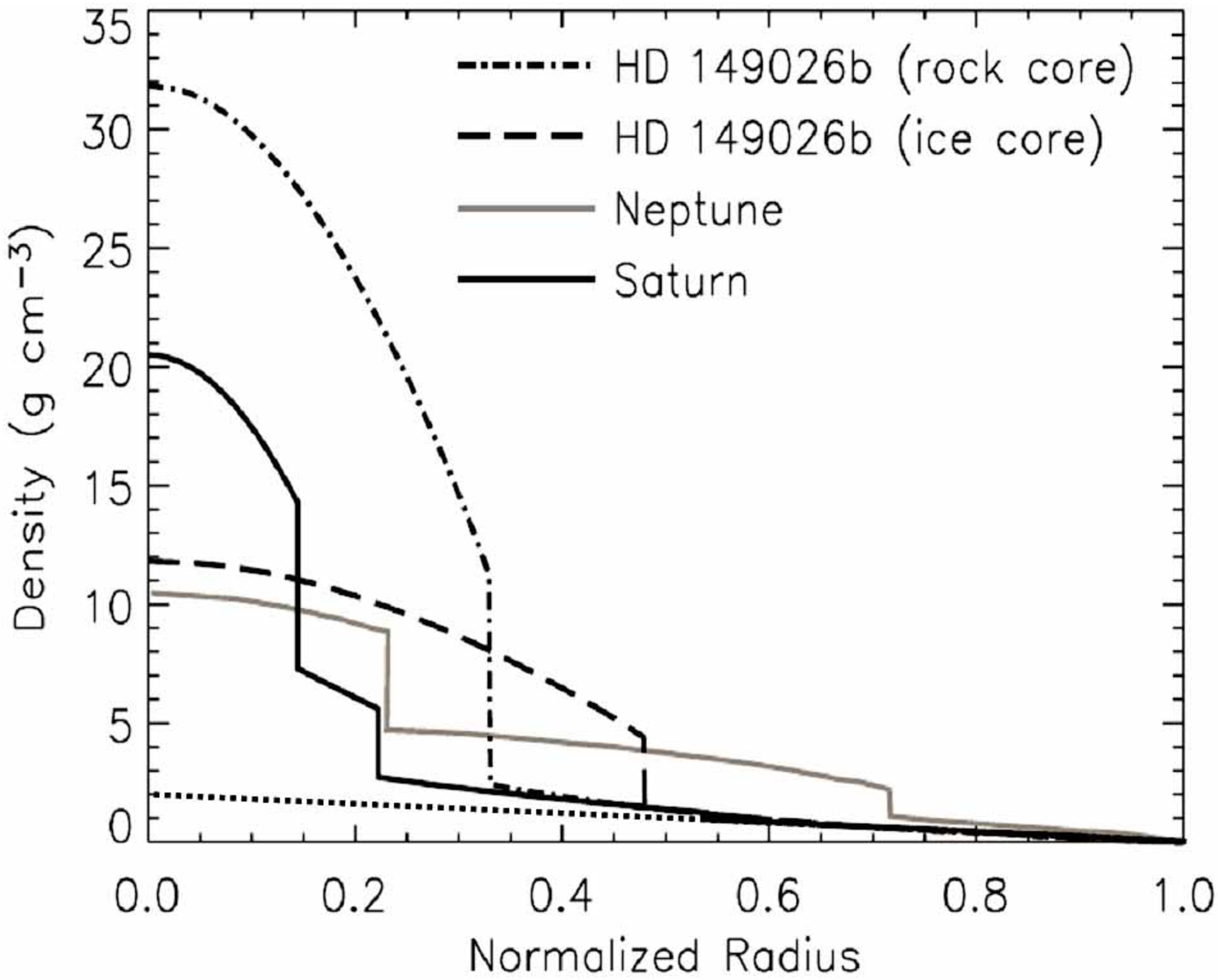}
\caption{\footnotesize{A few  Gas Giant Planets density profile;
the external saturnian density curve maybe approximated (dotted
line), for nearly half its radius, by a simple linear law: $\rho
\simeq (3.33 \cdot 10^{-5}\frac{h}{km} + 2\cdot 10^{-3} \,) g
cm^{-3}$ where h is the growing depth from the surface downward.
For these density profiles we derived the Tau length curves. See
Fig.\ref{fig3}. }} \label{fig6}
\end{figure}
\begin{figure}[t]
\includegraphics[width=70mm]{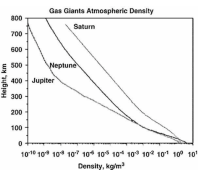}
\caption{\footnotesize{Main Giant Gaseous Planet density profile,}
\cite{Justus}} \label{fig7}
\end{figure}

\subsection{\textit{Inner Planets atmospheres}}
   The nearest planet as Mercury do not offer an atmosphere as
   the most distant ones. Therefore we consider for skimming UHECR and Tau showering in inner planets Mars
   and Venus. The Venus distance $d_{\tau}$ at horizons is wider by $40\%$ respect the
   Earth one: $d_{\tau }\simeq\sqrt{R\,h_0\pi/2}\simeq1.4\;d_{\tau\oplus}=504$ km corresponding to a maximal
    tau energy  $E_{\tau}\simeq[d_{\tau}/(c \tau_{\tau})]m_{\tau}c^2\simeq10$ EeV .
    On the other side, the Mars air density is too small ($\simeq1\%$ of earth one) to be of
    great interest for UHECR or EeV neutrinos induced showers. Moreover the Mars radius in nearly half
    of the terrestrial one, (but its height scale is a little longer):
    $d_{\tau }\simeq\sqrt{R\,h_0\pi/2}\simeq0.855\;d_{\tau\oplus}=308$ km, corresponding to a smaller
    tau energy $E_{\tau}\simeq6$ EeV, anyway undetectable by Mars thin atmosphere. Indeed only PeV or TeV
    cosmic rays might shower at maximal size at martian horizon.
    This possibility may  be used in gamma astronomy, if future telescope
    on   Mars will consider this kind of astronomy. Nevertheless such a low density atmosphere
    occurs already in high altitude balloons near Earth or
    in highest altitude for skimming gamma at top terrestrial altitudes.

\section{\textit{The Tau  length $l_{\tau}$ in Outer  Planets}}

 The parameter that must be correlated with the planet size is
  the propagation length $l_{\tau}$, see Fig. \ref{fig3}.
  The great advantage of largest gaseous  planets in UHE tau-showering is three fold:
  their size is wider, their height scale is longer and their
  external density profile are more diluted that terrestrial ones,
  offering in this way an extreme windows to maximal neutrino energies.

The nearest Jupiter planet, see Fig. \ref{fig4}, exhibit the
 largest radius and it surge as the leading planet for largest
 $l_{\tau}$. Indeed its radius, $R_J=71492$ km, is eleven
 times the Earth radius, while the atmosphere height $h_{0J}=3.4\,h_{0\oplus}$.
 Therefore the consequent maximal horizontal distance
is
$d_{J\tau}\simeq\sqrt{R_J\,h_{0J}\pi/2}\simeq6.15\;d_{\tau\oplus}=2214$
km, while $h_{0J}\simeq3.37h_{0\oplus}$, corresponding to a
maximal tau energy
$E_{J\tau}\simeq[d_{J\tau}/(c\tau_{\tau})]m_{\tau}c^2\simeq4.43\cdot10^{19}$
eV.

Surprisingly the best planet for largest tau distances is not
Jupiter but the nearby Saturn  Fig. \ref{fig5}: the radius $R_S$
is slightly smaller, but its height atmosphere scale is quite
larger. Indeed the Saturn radius  $R_S = 60268$ km, is only $9.4$
times our planet radius, but the atmosphere height scale $h_{0S}$
(see dotted line in  Fig. \ref{fig6} applied in Fig. \ref{fig3}),
as better compared in density profile Fig. \ref{fig7} ,
\cite{Justus}, is $7.4$ times the terrestrial $h_{0 \oplus}$.
Therefore the resulting
$d_{S\tau}\simeq\sqrt{R_S\,h_{0S}\pi/2}\simeq8.38\;d_{\tau\oplus}=3017$
km, is the largest in our planetary system. Then the maximal
energy for tau is
$E_{S\tau}\simeq[d_{S\tau}/(c\tau_{\tau})]m_{\tau}c^2\simeq6\cdot10^{19}$
eV. As it is shown in Fig. \ref{fig3}, where Saturn size distance
/dot-dashed line) intersect with the tau boosted fly distance. The
case of Uranus and Neptune planets are comparable (but of less
interest) because of their smaller sizes and smaller height
density growth, as well as because of their larger distances from
us. Moons around planets are usually with null atmosphere except
for a Saturnian moon, Titan, of great and peculiar atmosphere
density.
\subsection{\textit{The Titan Role in UpTaus showering}}
\begin{figure}[t]
\includegraphics[width=70mm]{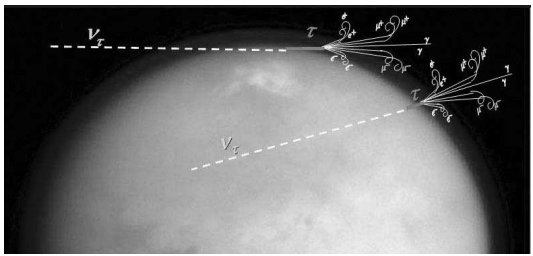}
\caption{\footnotesize{The Titan atmosphere in the edge and the
HorTaus and UpTaus showering along the smaller moon size} }
\label{fig8}
\end{figure}
\begin{figure}[t]
\includegraphics[width=70mm]{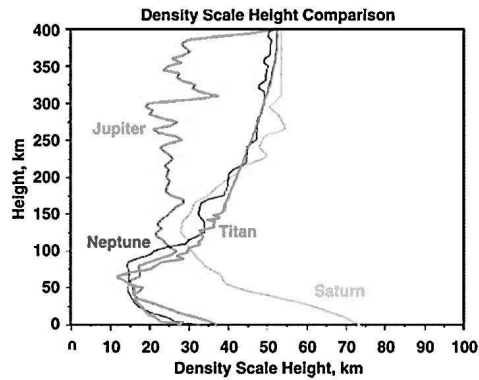}
\caption{\footnotesize{The density height scale for the main giant
gaseous planets} \cite{Justus}} \label{fig9}
\end{figure}
 One of the critical role of Tau Air-Showers is the parental
 UHE $\nu_{\tau}$ opacity crossing along the planet cord.
 The opacity is related to the electro-weak cross-section dependence
 on incoming neutrino energy with matter \cite{Gandhi98} and, of course, on the planet size and composition.
 Smallest radius allows greater neutrino energy: for the Earth
 diameter the neutrino cut-off is about $4\cdot10^{13}$ eV for $\nu_{\mu}$ and
$10^{15}$ eV for $\nu_{\tau}$ (because of their marginal
regeneration and  pile up by higher energies neutrinos toward PeV
band). For this reasons Horizontal Tau at tens PeV or EeV are
possible as well as up-taus just at PeV energies on Earth (or with
a partial suppression \cite{Fargion2004}).

\begin{figure}[t]
\includegraphics[width=70mm]{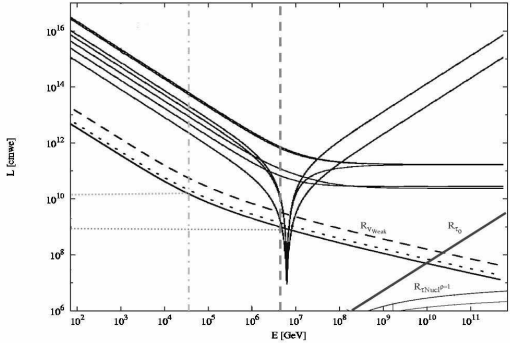}
\caption{\footnotesize{The interaction length for UHE neutrinos as
a function of their energy, crossing the Earth (green dot-dashed
line) and Titan (red dashed line)}} \label{fig10}
\end{figure}

Naturally Titan being just $5150$ km size and $1.88$ average
density allows higher energy neutrino crossing, because in
vertical direction (slant depth $X_{max}=9.68\cdot10^8$ w.e.) the
out-coming energy (see Fig.\ref{fig8}) maybe unsuppressed even a
hundred times at higher values: $E_{\tau min}\simeq5\cdot10^{15}$,
see Fig.\ref{fig10}, in principle a much better screen to look for
Up-Taus showering in the future search of UHE $\nu_{\tau}$ in
space. Finally up-ward and inclined tau air-shower at Titan may
enjoy of  the longer atmosphere height scale (respect to Earth),
which is $h_{0\,Titan}\simeq30$ km, see Fig \ref{fig9}, leading to
better contained upward showering event at $10^{17}-10^{18}$
\begin{figure}[h]
\includegraphics[width=70mm]{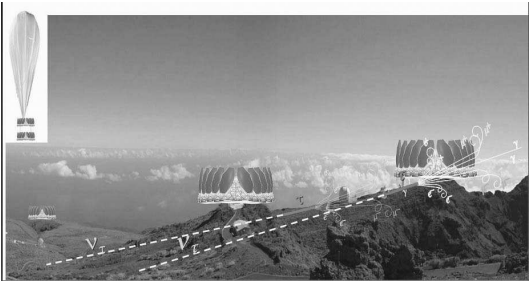}
\caption{\footnotesize{Ideal arrays of Cherenkov Crowns Telescopes
in Canaries \cite{FarCrown} and an equivalent twin Crown Array
Balloon in flight; similar arrays maybe located
 in planes or satellites. }} \label{fig11}
\end{figure}

\section{Back to the Earth: Magic and Auger}
\begin{figure}[t]
\includegraphics[width=80mm]{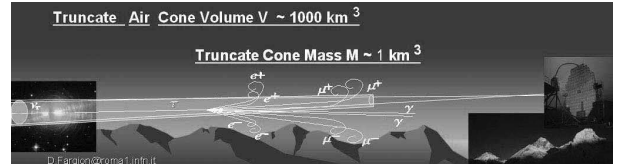}
\caption  {\footnotesize{The possible horizontal air-showering by
a GRB or an active BL Lac , whose UHE anti-electron neutrino might
resonance with air electrons at Glashow PeVs energies (or in
 Tau air-showers at higher energy), making  nearly $3\%$ of these GRB,SGRs,BL Lac
 sources  laying at horizons for Magic  Telescopes. The mass observed , as estimated in figure, within the
 air-cone exceed the $km^3$ water mass, even if within a narrow solid angle ($\simeq 4\cdot 10^{-3}$ sr.)}See \cite{Fargion2005}}
 \label{fig12}
\end{figure}
\begin{figure}[t]
\includegraphics[width=80mm]{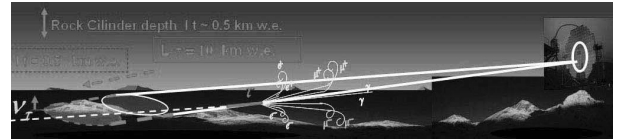}
 \caption  {\footnotesize{As above  EeVs tau  are originated in the Earth crust
 and while escaping the soil are testing $\sim 70-100 km^3$ volumes; later
 UHE tau may decay in flight and may air-shower loudly toward Magic telescope, within an area of
 few or tens $km^2$.} See \cite{Fargion2005}.}
 \label{fig13}
 \end{figure}
\begin{figure}[t]
\includegraphics[width=70mm]{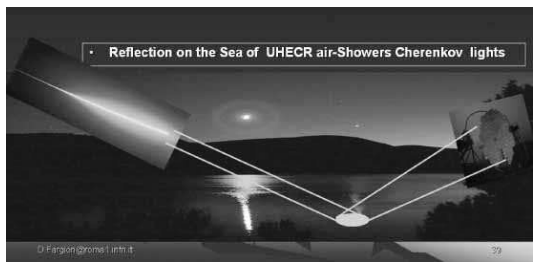}
 \caption  {\footnotesize{The possible inclined UHECR air-showering on
Magic facing the sea side. Their detection rate is large (at
zenith angle $80-85^o$) (tens or more a night)
 nearly comparable with those  at zenith angle  $87^o$ already estimated;
  these mirror UHECR shower , widely spread in oval
  images on the sea (depending on the sea wave surfaces), their presence
is an useful test for Magic discovering of point source PeV-EeV
UHECR air-showers at horizons.
 While previous configuration above horizons may correlate direct muon bundle and
Cherenkov flashes, these mirror events are polarized lights mostly
muon-free, diffused in large areas and dispersed in longer time
scales, mostly in twin (real-mirror-tail) spots. On the contrary
Up-going Tau air-showers from the sea are very beamed and thin and
un-polarized and brief.} See \cite{Fargion2005}.}
 \label{fig14}
\end{figure}
\begin{figure}[t]
\includegraphics[width=60mm]{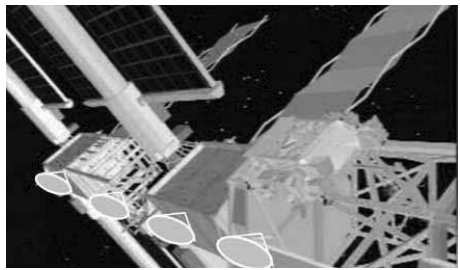}
 \caption  {\footnotesize{Array detector in Space facing the horizons \cite{Fargion2001}.} }
 \label{fig15}
\end{figure}
\begin{figure}[t]
\includegraphics[width=60mm]{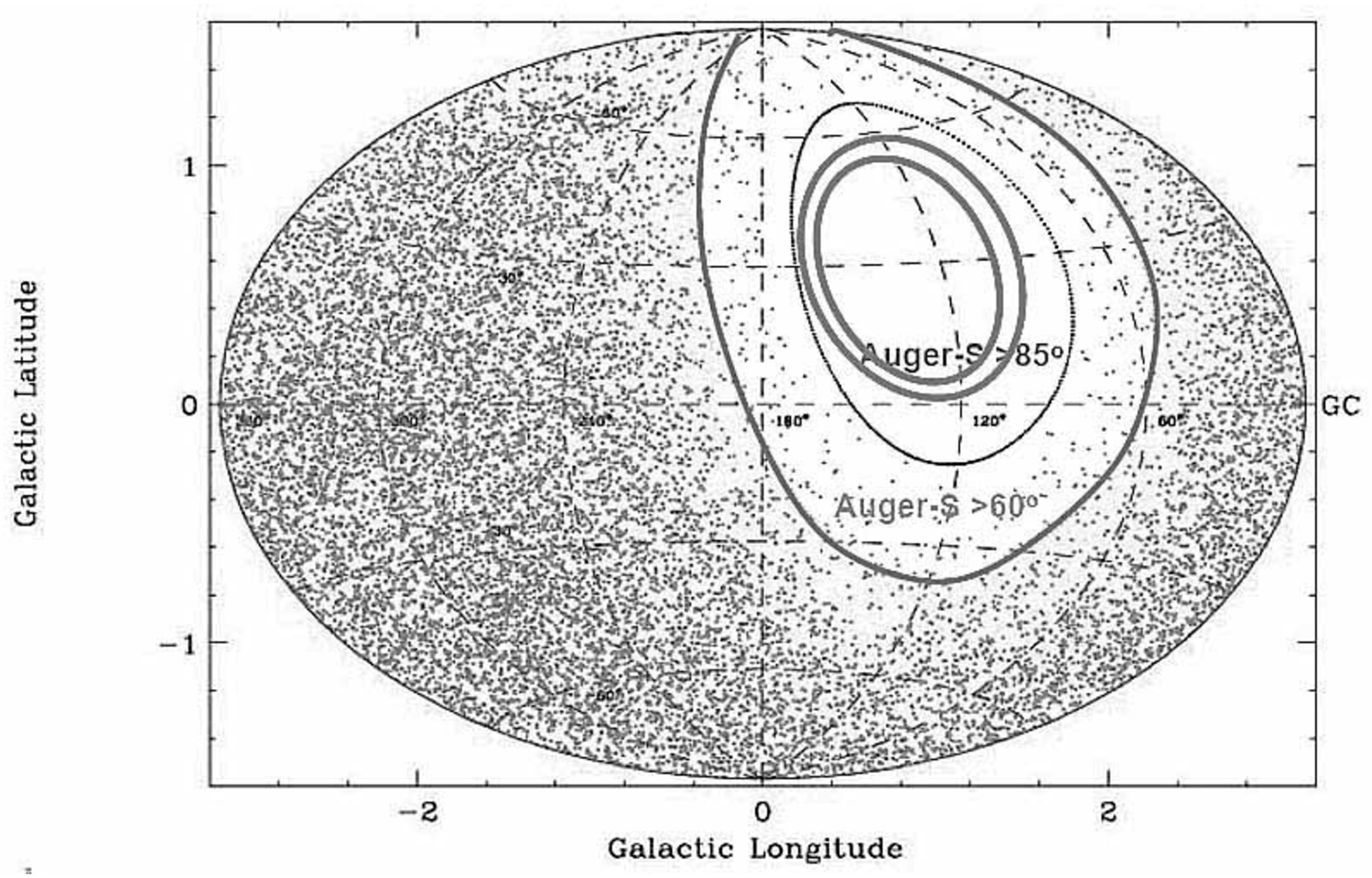}
\caption{\footnotesize{Description of the main Galactic Map of
AUGER and the Zenith areas $\geq 60^o$, $\geq 85^o$, and shadow
ring $\simeq 87^o-88^o$ where to find, within the Ande Shadows,
the eventual rare Young Horizontal Tau Air-Shower,
\cite{Fargion1999},\cite{Fargion 2002a},
\cite{Bertou2002},\cite{Cronin2004}, \cite{Fargion2004},
\cite{Fargion2004b},\cite{Miele et. all05} }} \label{fig16}
\end{figure}

  Naturally the simplest and nearest and most practical way to look for UHE neutrinos
 showering  at PeVs energies or above is facing the Sky of our Sky: the Earth.
 The possible way to trace such air-showers
 are deep valleys (to enhance the solid angle of the tau air-showers),
  peak mountains, balloons, planes and satellites facing the Earth.
 In UHE neutrino and $\tau$ showering search different experimental frame-work might be
  used: Cherenkov telescope and arrays, Scintillator and C.R.
  arrays  facing  Mountains or the Earth  edges See Fig.\ref{fig11}, \cite{FarCrown}.
Cherenkov gamma Telescopes as   MAGIC ones at the top of a
mountains are searching for tens GeV $\gamma$ astronomy. The same
telescope at zero cost in cloudy nights, may turn (for an bending
angle $\simeq10^{\circ}$) toward terrestrial horizontal edges,
testing both common PeVs cosmic ray air showers and UHE showering
in air, see Fig.\ref{fig12}. In upward directions muon and-or
gamma secondary  bundles by  up-going tau air-showers might blaze
the Telescope, see Fig.\ref{fig13}.  The peculiar Magic position
on the sea offer the geometry for  reflected downward CR  on
 the water. The absence of correlated muon bundle and the presence of
 a polarization in Cherenkov lights make a clear signature of these
 mirrored events. Indeed the possible detection of a far
 (un-mirrored) air shower is enriched by:
$1)$ early Cerenkov flash even if dimmed by atmosphere screen,
$2)$ single and multiple muon bundle shining Cerenkov rings or
arcs inside the Magic disk  in time correlation, $3)$ muon
decaying into electromagnetic in flight making mini showers mostly
outside the disk leading to lateral correlated gamma tails. We
estimated the rate for such PeVs-EeV events each night, finding
hundreds event of noises muons and tens of bundle correlated
signals each night \cite{Fargion2005}, \cite{Fargion2006} at
horizons. Among them up-going Tau Air-Showers may occur very
rarely, but their discover is at hand for dedicated $360^{\circ}$
crown Arrays on Mountains (see Fig.\ref{fig11}) \cite{FarCrown}or
in Space (see Fig.\ref{fig15}) in correlation among Cherenkov  and
additional scintillator detectors. The UHECR and UHE neutrino
astronomy at horizons may test  the expected  GZK \cite{za66}
neutrino traces and  eventual Z- Burst UHE neutrino parental
spectra at highest energy edges \cite{Fargion-Mele-Salis99},
\cite{Yoshida1998}, \cite{Quigg}.

\section{Conclusions}
The search for UHECR and UHE neutrino Astronomy is compelling: the
puzzles they may solve are fundamental. While UHE neutrino
telescope underground are still seeking rare single muon tracks,
UHE tau air-shower Astronomy may amplify the neutrino and the
UHECR skimming the Earth or the top atmosphere. The tau lenght is
related to the UHE neutrino energy and to allowable air layer on
Earth. New opportunities arise in outer planets as Jupiter and
Saturn, whose radiuses and whose height scale afford huge
distances and largest tau energies. A very peculiar case arise
around the small, but denser Titan atmosphere able to permit
up-tau at energies (tens PeV) very favorable for UHE neutrino
astronomy. However just to begin the Earth from mountains , in
deep valleys  and on balloons and space is the most actual place
to play the search for this novel Astronomy. The Auger shadows
from the Ande might hide a couple of event each year while Magic
Telescope might rush in Horizons edges at GRB event. Even
satellites like Pamela and Glast might test up-going showers as
Terrestrial Gamma Flashes whose nature might be already associated
to  UHECR or Tau showering skimming the Earth.

\end{document}